\documentclass[11pt,a4paper]{article}
\usepackage{fullpage}

\usepackage{amssymb}
\usepackage{color}

\usepackage{graphicx}
\usepackage{epstopdf}
\usepackage{caption}
\usepackage{subcaption}
\usepackage{booktabs}

\usepackage{url}
\urldef{\mailsm}\path|suejb.memeti@lnu.se|
\urldef{\mailll}\path|lu.li@liu.se|
\urldef{\mailsp}\path|sabri.pllana@lnu.se|
\urldef{\mailjk}\path|jokolodziej@pk.edu.pl|
\urldef{\mailck}\path|christoph.kessler@liu.se|

\usepackage{pbox}
\usepackage{array}

	\title{Benchmarking OpenCL, OpenACC, OpenMP, and CUDA: programming productivity, performance, and energy consumption}
	
	\author{Suejb Memeti\thanks{Linnaeus University, 351 95 V\"{a}xj\"{o}, Sweden, \mailsm}
		\and Lu Li\thanks{Link\"{o}ping University, 581 83 Link\"{o}ping, Sweden, \mailll}
		\and Sabri Pllana\thanks{Linnaeus University, 351 95 V\"{a}xj\"{o}, Sweden, \mailsp}
		\and Joanna Ko\l{}odziej\thanks{Research and Academic Computer Network (NASK), 01 045 Warsaw, Poland, \mailjk}
		\and Christoph Kessler\thanks{Link\"{o}ping University, 581 83 Link\"{o}ping, Sweden, \mailck}
		}

\date{\emph{Preprint}} 

\makeatletter
\let\@fnsymbol\@arabic
\makeatother

\begin{document}
	
	\maketitle
	
	\begin{abstract}
		Many modern parallel computing systems are heterogeneous at their node level. Such nodes may comprise general purpose CPUs and accelerators (such as, GPU, or Intel Xeon Phi) that provide high performance with suitable energy-consumption characteristics. However, exploiting the available performance of heterogeneous architectures may be challenging. There are various parallel programming frameworks (such as, OpenMP, OpenCL, OpenACC, CUDA) and selecting the one that is suitable for a target context is not straightforward. 
		In this paper, we study empirically the characteristics of OpenMP, OpenACC, OpenCL, and CUDA with respect to programming productivity, performance, and energy. To evaluate the programming productivity we use our homegrown tool CodeStat, which enables us to determine the percentage of code lines that was required to parallelize the code using a specific framework. We use our tool x-MeterPU to evaluate the energy consumption and the performance. Experiments are conducted using the industry-standard SPEC benchmark suite and the Rodinia benchmark suite for accelerated computing on heterogeneous systems that combine Intel Xeon E5 Processors with a GPU accelerator or an Intel Xeon Phi co-processor.
	\end{abstract}
	
	\section{Introduction}
	
	Modern parallel computing systems may comprise multi-core and many-core processors at their node level. Multi-core processors may have two or more cores, and usually run at a higher frequency than many-core processors. While multi-core processors are suitable for general-purpose tasks, many-core processors (such as the Intel Xeon Phi \cite{chrysos2014intel} or GPU \cite{gpu}) comprise a larger number of lower frequency cores that perform well on specific tasks.
	
	Due to their different characteristics, engineers often combine multi-core and many-core processors to create the so-called heterogeneous nodes that may, if carefully utilized, result in high performance and energy efficiency. Yan et al. \cite{yancomparison} highlight the importance of efficient node-level execution of parallel programs also for future large-scale computing systems \cite{Abraham2015}. However, utilizing the available resources of these systems to the highest possible extent require advanced knowledge of vastly different parallel computing architectures and programming frameworks \cite{Viebke2015,Memeti2015}.
	
	Some of the widely used parallel programming frameworks for heterogeneous systems include OpenACC \cite{Wienke:2012}, OpenCL \cite{opencl2010stone}, OpenMP \cite{openmp2013}, and NVIDIA CUDA \cite{cuda}. The challenge for the program developer is to choose one of the many available parallel programming frameworks that fulfills in the specific context the goals with respect to programming productivity, performance, and energy consumption. Existing work has systematically studied the literature about GPU-accelerated systems \cite{mittal2015survey}, compared programming productivity of OpenACC and CUDA using a group of students to parallelize sequential codes \cite{accandcuda2016}, or used kernels for comparing execution time of OpenCL and CUDA-based implementations.
	
	In this paper, we benchmark four well-known programming frameworks for heterogeneous systems: OpenMP, OpenACC, OpenCL, and CUDA. In addition to the industry-standard benchmark suite SPEC Accel \cite{juckeland2014spec}, we use the popular Rodinia \cite{che2009rodinia} benchmark suite to evaluate programing productivity, energy efficiency, and performance. We use our tool developed for this study \emph{CodeStat} to quantify the programming effort for parallelizing benchmark suites under study. Furthermore, we developed \emph{x-MeterPU} that is an extension of MeterPU, which enables us to measure the performance and energy consumption on systems that are accelerated with the Intel Xeon Phi and GPU. We present and discuss results obtained on two heterogeneous computing systems: \emph{Emil} that comprises two Intel Xeon Phi E5 processors and one Intel Xeon Phi co-processor, and \emph{Ida} that has two Intel Xeon Phi E5 processors and one GTX Titan X GPU. 
	
	The rest of this paper is structured as follows. Section \ref{sec:method} describes our approach, tools and infrastructure for benchmarking programs developed with OpenMP, OpenCL, OpenACC, and CUDA. Results of the experimental evaluation using SPEC and Rodinia benchmark suites are described in Section \ref{sec:evaluation}. Section \ref{sec:rw} discusses the related work. We conclude the paper in Section \ref{sec:summary}.
	
	\section{Our Methodology and Tools}
	\label{sec:method}
	
\begin{figure}
		\centering
		\includegraphics[width=0.5\textwidth]{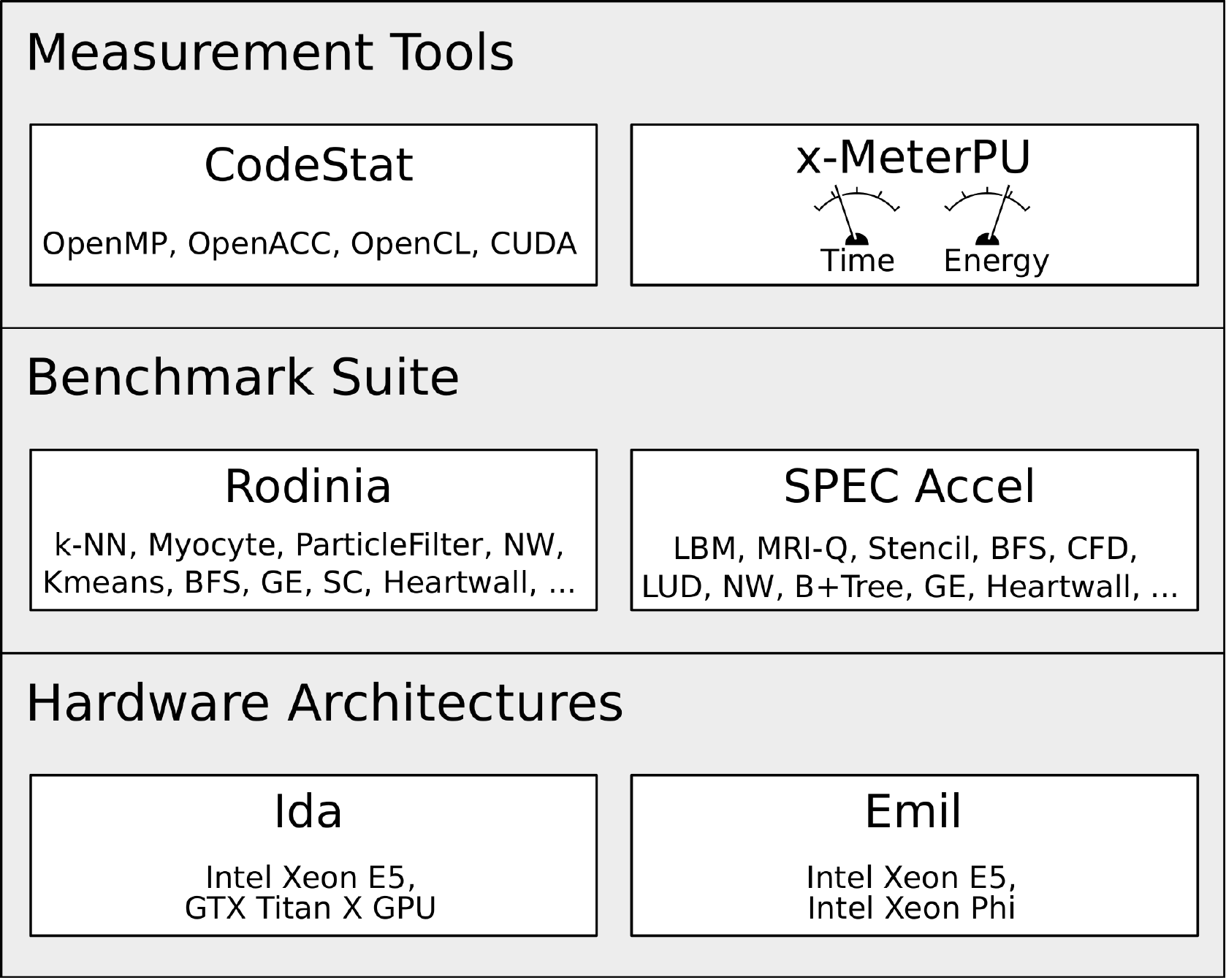}
		\caption{An overview of our infrastructure for evaluation of parallel programming languages for accelerated systems.}
		\label{fig:method}
	\end{figure}

	In this section, we describe our approach and tools that we have developed for benchmarking parallel frameworks for heterogeneous systems. Our approach is based on using industry-standard benchmark suites for evaluation of parallelization effort, performance, and energy consumption. Figure \ref{fig:method} depicts our infrastructure for evaluation of parallel programming languages for accelerated systems that includes measurement tools CodeStat and x-MeterPU, benchmark suites Rodinia and SPEC Accel, and heterogeneous systems Ida and Emil.
	
	Table \ref{table:frameworks-comparison} summarizes major features of parallel languages that are considered in our study: OpenMP \cite{openmp2013}, OpenACC \cite{Wienke:2012}, OpenCL \cite{opencl2010stone}, and CUDA \cite{cuda}. OpenMP and OpenACC are largely implemented as \emph{compiler directives} for C, C++, and FORTRAN, which significantly hide architecture details from the programmer. OpenCL and CUDA are implemented as software libraries for C and C++ and expose the programmer to low-level architectural details. With respect to the parallelism support, all of the considered frameworks support data parallelism and asynchronous task parallelism. While OpenMP and OpenCL provide parallelization patterns for both host multi-core CPUs and many-core accelerated devices, OpenACC and CUDA support parallelization means only for accelerators such as NVIDIA GPUs \cite{yancomparison}.
	
	\begin{table}[t]
    \footnotesize
		\centering
		\setlength{\tabcolsep}{4pt}
		\caption{Major features of OpenACC, OpenMP, OpenCL, and CUDA.}
		\label{table:frameworks-comparison}
		\begin{tabular}{>{\bfseries}p{2.7cm} p{2.7cm} p{2.7cm} p{2.7cm} p{2.7cm}}
			\toprule
			& OpenACC & OpenMP & OpenCL & CUDA \\ \toprule
			Parallelism & \pbox{2.4cm}{- data parallelism\\ - asynchronous task parallelism\\ - device only} & \pbox{2.4cm}{- data parallelism\\ - asynchronous task parallelism\\ - host and device} & \pbox{2.4cm}{- data parallelism\\ - asynchronous task parallelism\\ - host and device} & \pbox{2.4cm}{- data parallelism \\ - asynchronous task parallelism \\ - device only}     \\ \midrule
			
			\pbox{1.5cm}{Architecture \\abstraction} & \pbox{2.4cm}{- memory hierarchy\\ - explicit data mapping and movement} & \pbox{2.4cm}{- memory hierarchy - data and computation binding\\ - explicit data mapping and movement } & \pbox{2.4cm}{- memory hierarchy\\ - explicit data mapping and movement} & \pbox{2.4cm}{- memory hierarchy \\ - explicit data mapping and movement }      \\ \midrule
			
			\pbox{1.5cm}{Synchroni-\\zation} & \pbox{2.4cm}{- reduction\\ - join} & \pbox{2.4cm}{- barrier\\ - reduction\\ - join} & \pbox{2.4cm}{- barrier;\\- reduction} & \pbox{2.4cm}{- barrier }      \\ \midrule
			Framework implementation & compiler directives for C/C++ and Fortran & compiler directives for C/C++ and Fortran & C/C++ extension &  C/C++ extension     \\ \bottomrule
		\end{tabular}
	\end{table}

	\subsection{CodeStat - A tool for quantifying the parallelization effort}
	\label{sec:codestat}	
	
	To quantify the programming effort required to parallelize a code, we have developed our tool, named \emph{CodeStat}. \emph{CodeStat} takes as input a configuration file and the source code. The configuration file contains a list of valid file extensions (such as, \emph{.c, .cpp, .cu, .cl, ...}), and a list of framework specific method calls, or compiler directives (such as, \emph{\#pragma omp, proc\_bind,} or \emph{omp\_set\_num\_threads} in OpenMP). 
	
	CodeStat analyzes the code by looking for framework specific code-statements provided in the configuration file, and provides as output the total number of lines of code (LOC) and the number of LOC written in OpenMP, OpenCL, OpenACC, or CUDA. 
	
	The configuration files for OpenMP, OpenCL, OpenACC, and CUDA are provided by \emph{CodeStat}. Other programming frameworks can be supported by simply creating a new configuration file and adding the framework specific code-statements to the list. The list of statements for a given programming language is usually provided in the corresponding documentation or reference guides. 
	
	\subsection{x-MeterPU - A tool for performance and energy consumption measurement}
	\label{sec:x-meterpu}
	
	To measure the execution time and the energy consumption of the GPU accelerated systems we use MeterPU \cite{Lu-meterpu-journal}. MeterPU is a C++ software package that implements its generic yet simple measurement API. MeterPU allows easy measurement for different metrics (e.g., time, energy) on different hardware components (e.g. CPU, DRAM, GPU). Utilizing the template and meta-programming features of C++, MeterPU's overhead is quite low and unnoticeable. It enables easy modeling and tuning for energy, besides it also allows for smooth transition for legacy tuning software (e.g. SkePU\cite{Lu-meterpu-journal}) from time optimization to energy optimization if some assumptions used for time modeling are not violated for energy modeling. 
	
	To measure the execution time and the energy consumption of the Intel Xeon Phi, we have developed our tool, named \emph{x-MeterPU}. x-MeterPU supports energy measurements for both native and offload programming models of Intel Xeon Phi. It is able to automatically detect the execution environment, therefore a single API function is used for measurements for both native and offload applications. To measure the energy of offload-based applications, we use the \textit{micsmc} utility, whereas the \textit{micras} tool is used to measure the energy of the native-based applications. 
	
	Similarly to MeterPU, the use of x-MeterPU is very simple. The \textit{start()} and \textit{stop()} methods are used to enclose code regions to be measured. The \textit{get\_value()} function is used to retrieve the energy consumption (in Jules). In addition to the total energy consumption, x-MeterPU returns a log file containing all the power data with exact timestamps, which enables the production of various plots.
	
	\section{Evaluation}
	\label{sec:evaluation}
	
	In this section, we experimentally evaluate the selected parallel programming frameworks using various benchmark applications and architectures. We describe:
	
	\begin{itemize}
		\item the experimentation environment, including hardware configuration, benchmark applications, and evaluation metrics;
		\item the comparison results of OpenMP, OpenACC, OpenCL, and CUDA with respect to programming productivity, performance, and energy consumption.
	\end{itemize}

	\subsection{Experimentation Environment}
	
	In this section, we describe the experimentation environment used to evaluate the selected parallel programming frameworks on heterogeneous systems. We describe the hardware configuration, the considered benchmark applications, and the evaluation metrics.
	
	\subsubsection{Hardware Configuration}
	
	For experimental evaluation of the selected parallel programming frameworks, we use two heterogeneous single-node systems. 
	
	\textit{Emil} is a heterogeneous system that consists of two Intel Xeon E5-2695 v2 general purpose CPUs on the host, and one Intel Xeon Phi 7120P co-processing device. In total, the host is composed of 24 cores, each CPU has 12 cores that support two threads per core (known as logical cores) that amounts to a total of 48 threads. The Xeon Phi device has 61 cores running at 1.2 GHz base frequency, four hardware threads per core, which amounts to a total of 244 threads. One of the cores is used by the lightweight Linux operating system installed on the device. 
	
	\textit{Ida} is a heterogeneous system that consists of two Intel Xeon E5-2650 v4 general purpose CPUs on the host, and one GeForce GTX Titan X GPU. Similar to \textit{Emil}, \textit{Ida} has 24 cores and 48 threads on the host, whereas the GPU device has 24 Streaming Multiprocessors (SM), and in total 3072 CUDA cores running at base frequency of 1 GHz. The major features of \textit{Emil} and \textit{Ida} are listed in table \ref{table:system-configurations}.
	
	\begin{table}[t]
		\footnotesize 
		\centering
		\caption{The system configuration details for \textit{Emil} and \textit{Ida}.}
		\label{table:system-configurations}
		\begin{tabular}{@{}p{2.5cm}p{2.3cm}p{2.3cm}|p{2.3cm}p{2.3cm}@{}}
			\toprule
			& \multicolumn{2}{c}{Ida}		& \multicolumn{2}{c}{Emil}        \\ \midrule
			Specs  		   & Intel Xeon E5 & GeForce GPU	& Intel Xeon E5      & Intel Xeon Phi      \\ \midrule
			Type           & E5-2650 v4    & GTX Titan X	& E5-2695 v2    		& 7120P               \\
			Core Frequency & 2.2 - 2.9 GHz & 1 - 1.1 GHz	& 2.4 - 3.2 GHz 		& 1.2 - 1.3 GHz    \\
			\# of Cores    & 12            & 3072 			& 12            		& 61              	\\
			\# of Threads  & 24            & /          	& 24            		& 244                 \\
			Cache          & 30 MB         & /          	& 30 MB         		& 30.5 MB             \\
			Mem. Bandwidth & 76.8 GB/s     & 336.5 GB/s 	& 59.7 GB/s     		& 352 GB/s            \\
			Memory         & 384 GB        & 12 GB      	& 128 GB        		& 16 GB               \\
			TDP            & 105 W         & 250 W      	& 115 W         		& 300 W               \\ \bottomrule
		\end{tabular}
	\end{table}
	
	\subsubsection{Benchmark Applications}
	
	In this paper we have considered a number of different applications from the SPEC Accel and Rodinia benchmark suites. 
	The Standard Performance Evaluation Corporation (SPEC) Accel benchmark suite focuses on the performance of compute intensive parallel computing applications using accelerated systems. It provides a number of applications for OpenCL and OpenACC.
	Similarly, the Rodinia benchmark suite provides a number of different applications for OpenMP, OpenCL, and CUDA. 
	
	While SPEC Accel provides in total 19 OpenCL and 15 OpenACC applications, we have selected only 14 OpenCL and 3 OpenACC applications, whereas Rodinia provides 25 applications, however we have selected only 19 of them to use for experimentation (see table \ref{table:benchmark-apps}). The inclusion criteria during the selection process of applications are: (1) the need to have at least two implementations of the same application in different programming frameworks, or benchmark suites, and (2) applications that are compilable in our systems. For performance comparison of the selected parallel programming frameworks, we have used the following benchmark applications from the SPEC Accel and Rodinia benchmark suite: 
	
	\begin{itemize}
		
		\item \textbf{Lattice-Boltzmann Method (LBM)} - is a partial differential equation (PDE) solver in fluid dynamics. The SPEC Accel benchmark provides OpenCL and OpenACC implementation of this application.
		
		\item \textbf{MRI-Q} - is an application that is used for MRI construction in medicine. OpenCL and OpenACC implementations of this application are provided by SPEC Accel benchmark.
		
		\item \textbf{3D-Stencil} - a stencil code that represents the Jacobi iterative PDE solver in thermodynamics. This application is implemented using OpenCL and OpenACC as part of the SPEC benchmark, and the CUDA implementation as part of the Rodinia Benchmark. Unfortunately, there is no OpenMP implementation.
		
		\item \textbf{Breadth-First-Search (BFS)} - is a graph traversal problem, which is commonly used to find the shortest path between two nodes.
		
		\item \textbf{Computational Fluid Dynamics Solver (CFD)} - a solver for the three-dimensional Euler equations used in fluid dynamics.
		
		\item \textbf{HotSpot} - is a known application in physics simulation for thermal estimation of the processor based on the architecture and simulated power measurements.
		
		\item \textbf{LU decomposition (LUD)} - is a parallel application for calculating a set of linear equations.
		
		\item \textbf{Needleman-Wunsch (NW)} - application is used in bioinformatics, which is a non-linear optimization approach for alignments of DNA sequences. 
		
		\item \textbf{B+ Tree} - is an application that traverses B+trees. B+Tree represents sorted data that allows efficient insertion and removal of graph elements.
		
		\item \textbf{Gaussian Elimination (GE)} - is a dense linear algebra algorithm that solves all the variables in a linear system by computing the result row by row.
		
		\item \textbf{Heart Wall} - is another medical imaging application that is used for tracking the movement of the mouse heart over numerous ultrasound images. 
		
		\item \textbf{Kmeans} - is a data-mining application which is used for clustering based dividing the cluster on sub-cluster and calculating the mean values of each sub-cluster. 
		
		\item \textbf{LavaMD} - is a molecular dynamics application that calculates the potential and relocation of particles within a large 3D space.
		
		\item \textbf{Speckle Reducing Anisotropic Diffusion (SRAD)} - is an image processing algorithm based on PDEs for diffusing ultrasonic and radar imaging applications. 
		
		\item \textbf{Back Propagation (BP)} - is a machine-learning algorithm that is used during the training process of a layered neural network. 
		
		\item \textbf{k-Nearest Neighbor (k-NN)} - is a data-mining application that is used to find the k-nearest neighbors from an unstructured data set.
		
		\item \textbf{Myocyte} - is a simulation application used in medicine to model the cardiac myocyte (that is a heart muscle cell) and simulate its behavior.
		
		\item \textbf{Particle Filter (PF)} - is a medial imaging application that is used for tracking leukocytes and myocardial cells. However, this algorithm can be used in different domains, including video surveillance, and video compression. 
		
		\item \textbf{Streamcluster (SC)} - is a data-mining application that is used for on-line clustering of a given input stream. 
	\end{itemize}
	
	BFS, CFD, HotSpot, LUD, and NW are implemented using OpenCL OpenMP and CUDA. The OpenCL implementation is provided by both SPEC Accel and Rodinia benchmark suites. No OpenACC implementation is provided by SPEC Accel for these applications.
	B+Tree, GE, HW, Kmeans, LavaMD, and SRAD are implemented using OpenCL and CUDA. While the OpenCL implementation is provided by both benchmarks, the CUDA implementation is provided by Rodinia.
	BP, k-NN, Leukocyte, Myocyte, PathFinder, PF, and SC are implemented using OpenCL and CUDA, and their implementations are provided by the Rodinia benchmark suite.
	
	Additional information and implementation details for each of the considered benchmark applications are available at the documentation web-pages of SPEC Accel \cite{spec-docs} and Rodinia \cite{rodinia-wiki}.
	
	\begin{table}[t]
		\footnotesize
		\centering
		\caption{The considered applications from the SPEC Accel and the Rodinia benchmark suites.}
		\label{table:benchmark-apps}
		\begin{tabular}{@{}p{2cm}lcc|ccc@{}}
			\toprule
			& 			   & \multicolumn{2}{c}{SPEC Accel}     & \multicolumn{3}{c}{Rodinia}         \\ \midrule
			Application & 	Domain  			  & OpenCL       & OpenACC  & OpenMP    & OpenCL     & CUDA       \\ \midrule
			LBM 		&	Fluid Dynamics        & x       	 & x	    &           &            &            \\
			MRI-Q		&	Medicine              & x     		 & x	    &           &            &            \\
			Stencil		&	Thermodynamics        & x   		 & x 		&           &            &            \\
			
			BFS  		&	Graph Algorithms	  & x	         &          & x			& x			 & x		  \\
			CFD 		&	Fluid Dynamics        & x      		 &          & x         & x          & x          \\
			HotSpot 	&	Physics Simulation    & x			 &          & x         & x          & x          \\
			LUD 		&	Linear Algebra        & x 		     &          & x   		& x  	 	 & x	      \\
			NW 			&	Bioinformatics        & x   	     &          &    		& x 		 & x	      \\
			
			B+Tree 		&	Search                & x			 &          &           & x          & x     \\
			GE 			&	Linear Algebra        & x      		 &          &           & x			 & x		  \\
			Heartwall   &	Medical Imaging       & x			 &          &           & x          & x          \\
			Kmeans 		&	Data Mining           & x		     &          & 	        & x          & x          \\
			LavaMD 		&	Molecular Dynamics    & x		     &          & 	        & x          & x          \\
			SRAD 		&	Image Processing      & x 		     &          &           & x          & x          \\
			
			BP			&	Pattern Recognition   &              &          & 	        & x		     & x	      \\
			k-NN		&	Data Mining           &              &          & 		    & x			 & x		  \\
			Myocyte		&	Biological Simulation &              &          & 		    & x          & x          \\
			PF			&	Medical Imaging       &              &          & 		    & x		     & x	      \\
			SC			&	Data Mining           &              &          & 		    & x    		 & x	      \\
			
			\bottomrule
		\end{tabular}
	\end{table}

	\subsection{Evaluation Metrics}
	
	In this section, we discuss the evaluation metrics considered for comparison of the selected parallel programming frameworks, including the required programming effort, the performance, and the energy efficiency.
	
	\subsubsection{Programming Productivity}
	\label{sec:productivity}
	
	To quantitatively evaluate the programming effort required to parallelize a program, we use our tool named \emph{CodeStat} (see Section \ref{sec:codestat}). We use \emph{CodeStat} to determine the total lines of code $LOC_{total}$ and the fraction of lines of code $LOC_{par}$ written in OpenCL, OpenACC, OpenMP, or CUDA for a given application. We define the parallelization effort as follows,
	
	\begin{equation} \label{eq:productivity}
	Effort_{par} [\%] = 100 * LOC_{par}/ LOC_{total}
	\end{equation}
	
	\subsubsection{Performance and Energy Consumption}
	
	We use the execution time ($T$) and the Energy ($E$) to express the performance and energy consumption.	$T$ is defined as the total amount of time that an application needs from the start till the end of the execution, whereas $E$ is defined as the total amount of energy consumed by the system (including host CPUs and accelerators) from the beginning until the end of the execution. 
	
	The data for the execution time and the energy consumption are collected using the x-MeterPU tool (see Section \ref{sec:x-meterpu}). To collect such data, a wrapper class file was created. The control flow of this class is as follows: (1) start the time and energy counters; (2) synchronously execute the benchmark commands; and (3) stop the time and energy counters and calculate the total execution time and system energy consumption. 
	
	\subsection{Results}
	
	In this section, we compare OpenMP, OpenACC, OpenCL, and CUDA with respect to (1) programming productivity, and (2) performance and energy consumption. 
	
	\subsubsection{Programming Productivity}
	
	Table \ref{table:programmability} shows the parallelization effort as percentage of code lines written in OpenCL, OpenACC, OpenMP, or CUDA that are used to parallelize various applications of Rodinia and SPEC Accel benchmark suites. We use Equation \ref{eq:productivity} in Section \ref{sec:productivity} to calculate the percentage of code lines. 
	
	\begin{table}[t]
    \footnotesize
		\centering
		\caption{Programming effort required to parallelize the code is expressed as fraction of code lines written in OpenCL, OpenACC, OpenMP, or CUDA. The remaining code lines are written in general-purpose C/C++.}
		\label{table:programmability}
		\begin{tabular}{@{}lrr|rrr@{}}
			\toprule
			& \multicolumn{2}{c}{\textbf{SPEC Accel}}  & \multicolumn{3}{c}{\textbf{Rodinia}}              \\ \midrule
			\textbf{}           & \textbf{OpenCL[\%]} & \textbf{OpenACC[\%]} & \textbf{OpenMP[\%]} & \textbf{OpenCL[\%]} & \textbf{CUDA[\%]} \\ \midrule
			\textbf{LBM}        & 3.21  & 0.87 	&       &       &       \\
			\textbf{MRI-Q}      & 5.70  & 0.64 	&       &       &       \\
			\textbf{Stencil}    & 4.70  & 0.61 	&       &       &       \\
			\textbf{BFS}        & 6.95  &  		& 4.86	& 9.07  & 12.50	\\
			\textbf{CFD}        & 5.83  &  		& 2.53  & 9.00	& 8.08  \\
			\textbf{HotSpot}    & 4.75  &  		& 2.67  & 13.18 & 8.20  \\
			\textbf{LUD}        & 5.78  &  		& 2.30  & 9.72  & 7.82  \\
			\textbf{NW}         & 6.56  &  		&  		& 18.34 & 8.85  \\
			\textbf{B+Tree}     & 4.89  &  		&  		& 6.79  & 4.51  \\
			\textbf{GE}         & 9.63  &  		&  		& 14.21 & 9.76  \\
			\textbf{Heartwall}  & 5.34  &  		&  		& 6.74  & 3.97  \\
			\textbf{Kmeans}     & 2.80  &  		&  		& 2.67  & 2.17  \\
			\textbf{LavaMD}     & 4.61  &  		&  		& 9.24  & 7.74  \\
			\textbf{SRAD}       & 7.81  &  		&  		& 13.00 & 10.28 \\
			\textbf{BP}         &  	    &  		&  		& 12.21 & 5.95  \\
			\textbf{k-NN}       &  	    &  		&  		& 15.83 & 5.07  \\
			\textbf{Myocyte}    &  		&  		&  		& 8.25  & 1.21  \\
			\textbf{PF}         &  		&  		&  		& 17.83 & 9.47  \\
			\textbf{SC}         &  		&  		&  		& 5.81  & 2.66  \\ \bottomrule
		\end{tabular}
	\end{table}
	
	\textit{Result 1: Programming with OpenCL requires significantly more effort than programming with OpenACC for SPEC Accel benchmark suite.}
	
	Based on the available OpenCL and OpenACC code for \emph{LBM, MRI-Q}, and \emph{Stencil} from the SPEC Accel benchmark suite, we may observe that on average OpenACC requires about $6.7\times$ less programming effort compared to OpenCL.
	
	\textit{Result 2: Programming with OpenCL on average requires about two times more effort than programming with CUDA for Rodinia benchmark suite.}
	
	With respect to the comparison between OpenCL and CUDA using the applications in the Rodinia benchmark suite, except of the BFS implementation, on average CUDA requires $2\times$ less programming effort than OpenCL.
	
	\textit{Result 3: Programming with OpenMP requires less effort than programming with OpenCL and CUDA.}
	
	Based on the data collected for OpenMP, OpenCL, and CUDA implementations of \emph{BFS, CFD, HotSpot}, and \emph{LUD} from the Rodinia benchmark suite, we may observe that on average OpenMP requires 3.6$\times$ less programming effort compared to OpenCL, and about $3.1\times$ less programming effort compared to CUDA. 
	
	\textit{Result 4: The human factor can impact the fraction of code lines used to parallelize the code.}
	
For example, the OpenCL implementation of BFS on the SPEC Accel benchmark suite comprises 6.95\% OpenCL code lines, whereas the implementation on the Rodinia comprises 9.07\% OpenCL code lines. Differences in the parallelization effort can be observed also for \emph{CFD, HotSpot, LUD, NW, B+Tree, GE, Heartwall, Kmeans, LavaMD}, and \emph{SRAD}.	
	
	\subsubsection{Performance and Energy}
	
\begin{figure}
		\centering
		\begin{subfigure}[b]{1.04\textwidth}
			\includegraphics[width=\textwidth]{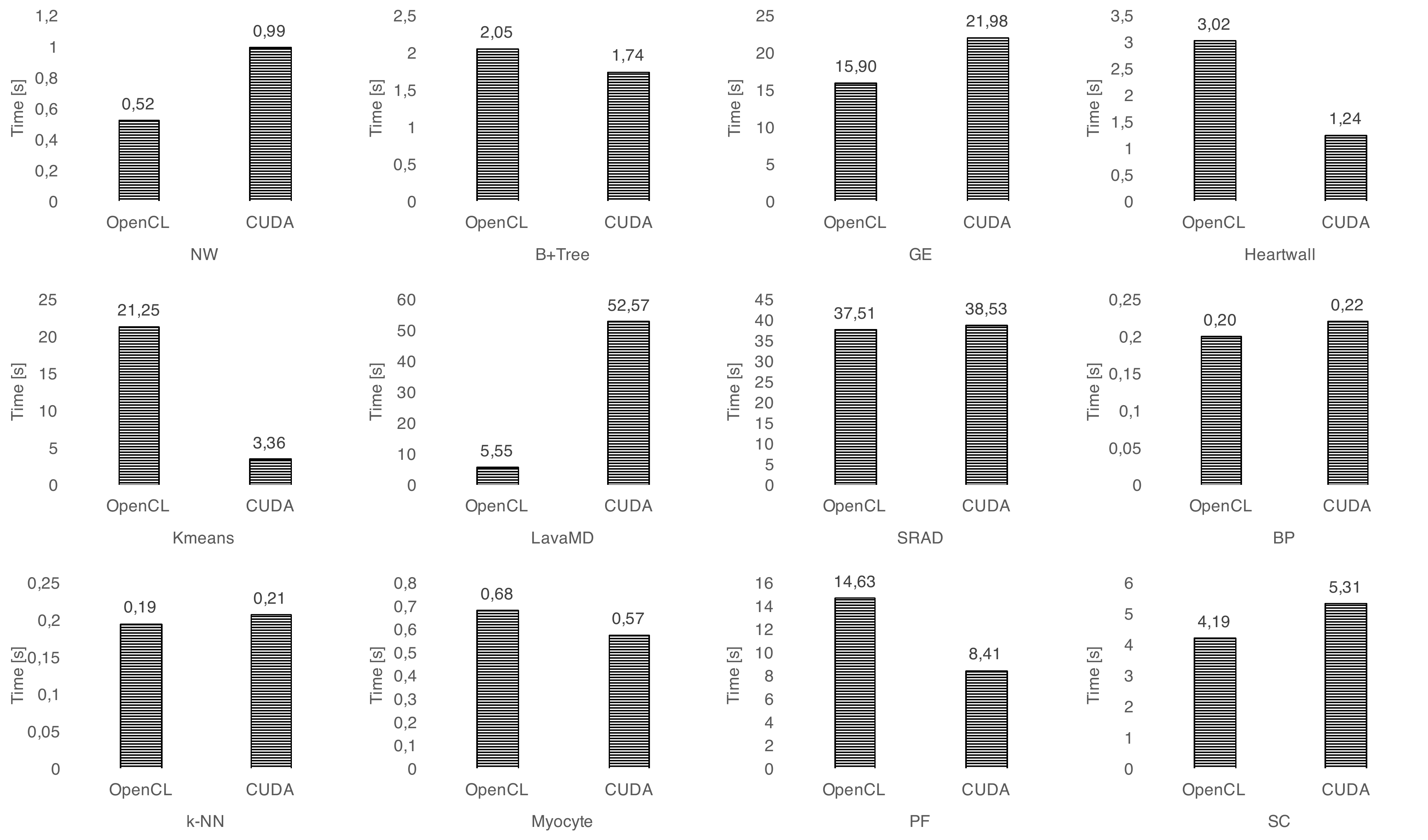}
			\caption{Time}
			\label{fig:time-rodinia-ida}
		\end{subfigure}
		~ 
		\begin{subfigure}[b]{1.04\textwidth}
			\includegraphics[width=\textwidth]{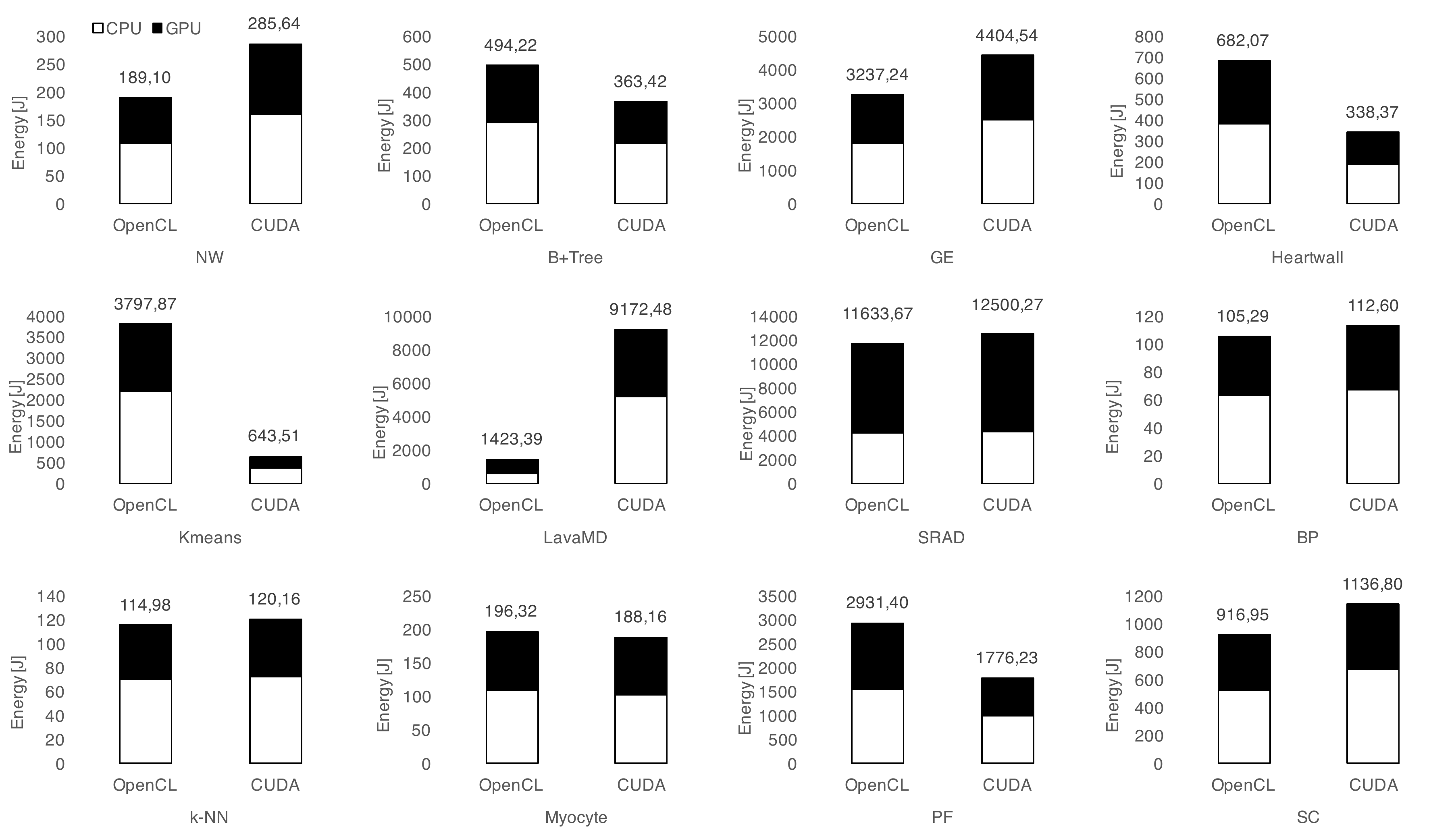}
			\caption{Energy Consumption}
			\label{fig:energy-rodinia-ida}
		\end{subfigure}
		\caption{A comparison of OpenCL and CUDA with respect to (a) execution time and (b) energy consumption. Results are obtained for various applications from the Rodinia benchmark suite (Table \ref{table:benchmark-apps}) on the GPU-accelerated system \emph{Ida} (Table \ref{table:system-configurations}).}
		\label{fig:time-energy-rodinia-ida}
	\end{figure}

\begin{figure}
		\centering
		\begin{subfigure}[b]{0.48\textwidth}
			\includegraphics[width=\textwidth]{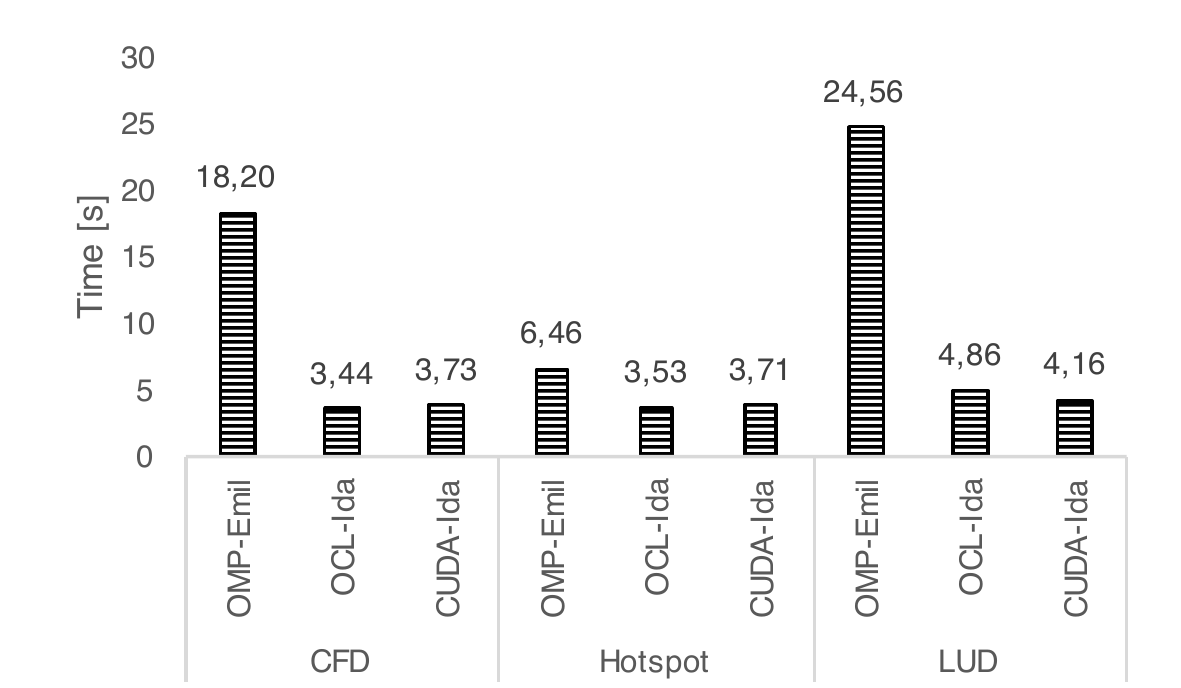}
			\caption{Time}
			\label{fig:time-cfd-hs-lud-rodinia-emil-ida}
		\end{subfigure}
		~ 
		\begin{subfigure}[b]{0.48\textwidth}
			\includegraphics[width=\textwidth]{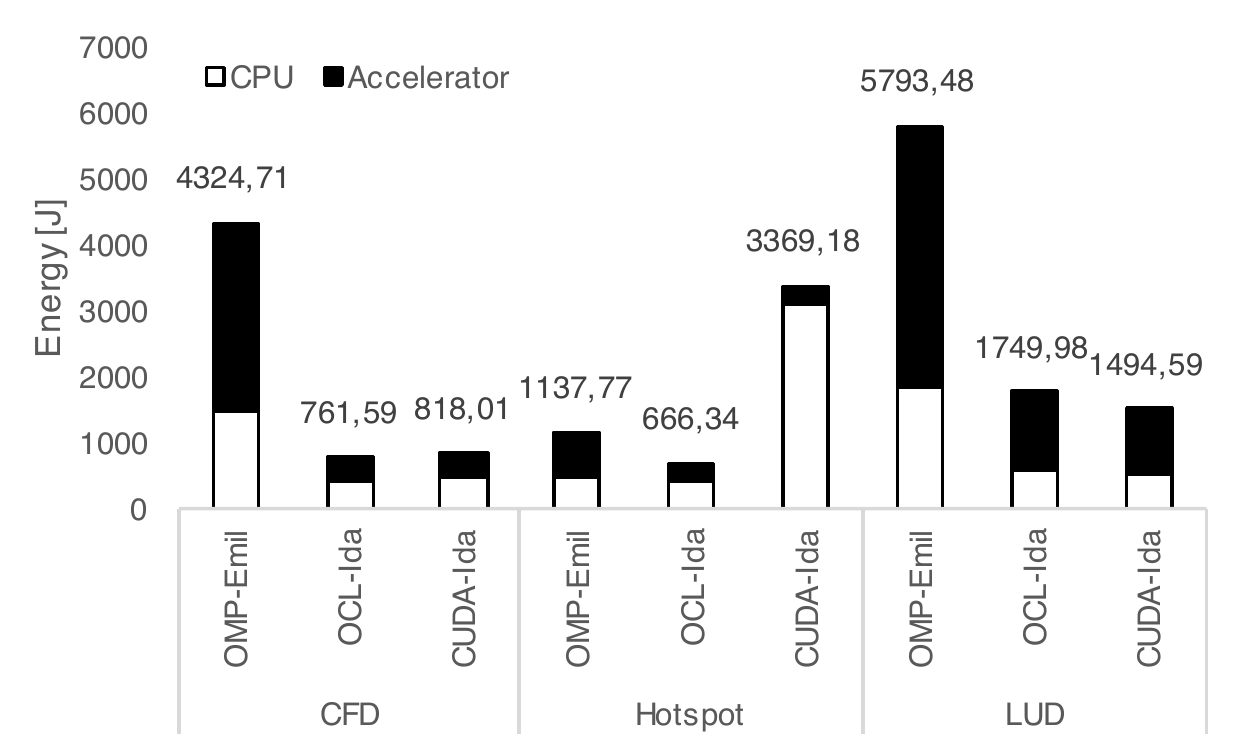}
			\caption{Energy Consumption}
			\label{fig:energy-cfd-hs-lud-rodinia-emil-ida}
		\end{subfigure}
		\caption{A comparison of OpenMP, OpenCL, and CUDA with respect to (a) execution time and (b) energy consumption using Rodinia benchmark suite (Table \ref{table:benchmark-apps}). OpenMP (OMP) version of CFD, HotSpot, LUD is executed  on the Intel Xeon Phi accelerated system \emph{Emil} (Table \ref{table:system-configurations}); whereas OpenCL (OCL) and CUDA versions of CFD, HotSpot, LUD are executed on the GPU-accelerated system \emph{Ida}.}
		\label{fig:time-energy-rodinia-emil-ida}
	\end{figure}

\begin{figure}
		\centering
		\begin{subfigure}[b]{0.48\textwidth}
			\includegraphics[width=\textwidth]{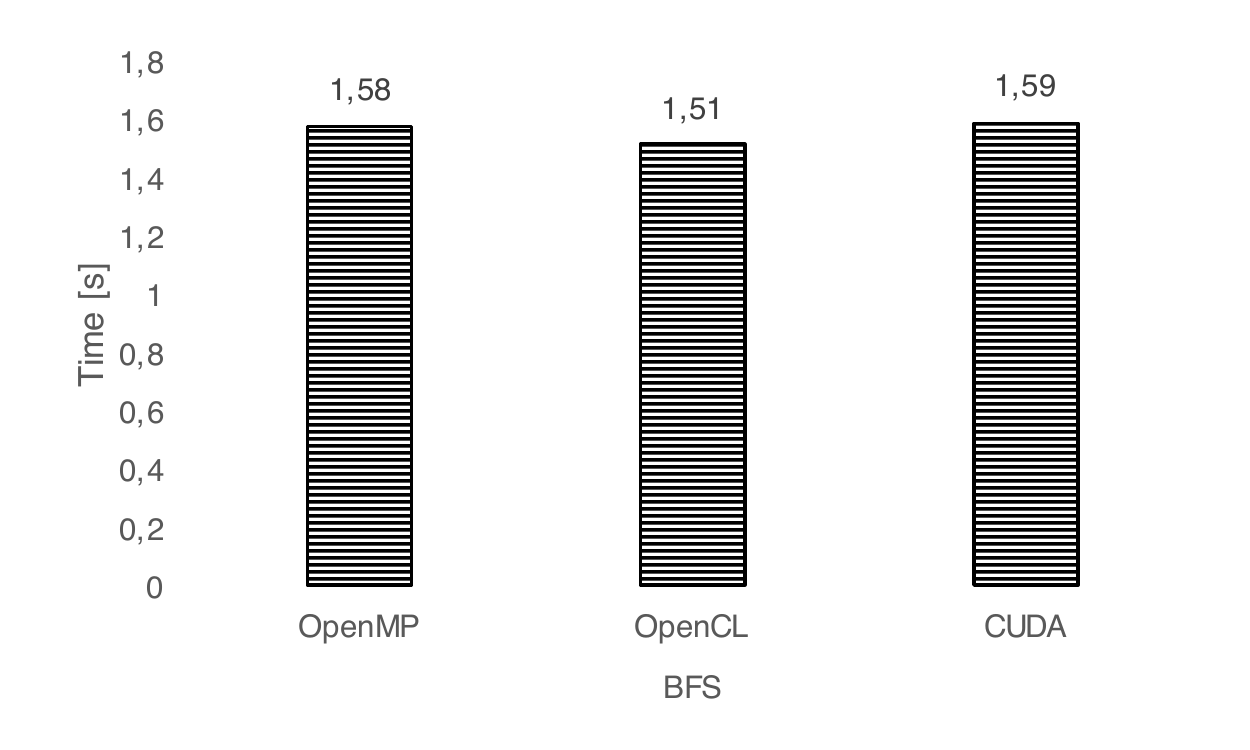}
			\caption{Time}
			\label{fig:time-bfs-rodinia-emil-ida}
		\end{subfigure}
		~ 
		\begin{subfigure}[b]{0.48\textwidth}
			\includegraphics[width=\textwidth]{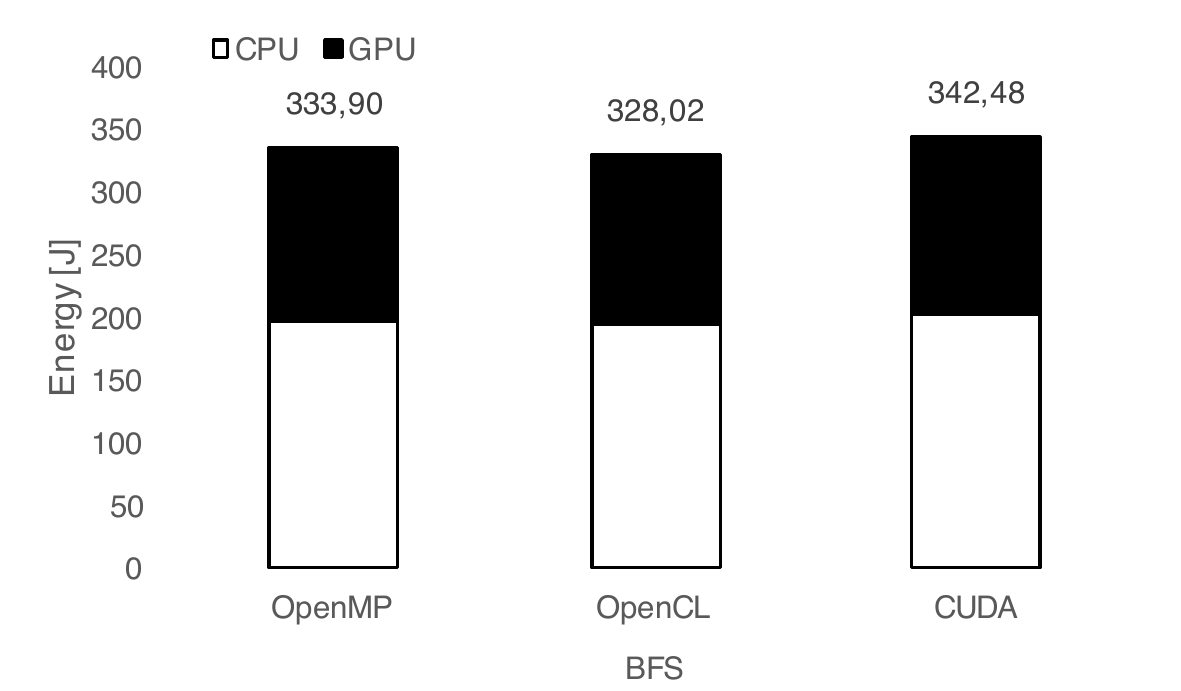}
			\caption{Energy Consumption}
			\label{fig:energy-bfs-rodinia-ida}
		\end{subfigure}
		\caption{A comparison of OpenMP, OpenCL, and CUDA with respect to (a) execution time and (b) energy consumption using BFS application from the Rodinia benchmark suite (Table \ref{table:benchmark-apps}) on \emph{Ida}.}
		\label{fig:time-energy-bfs-rodinia-ida}
	\end{figure}

\begin{figure}
		\centering
		\begin{subfigure}[b]{0.48\textwidth}
			\includegraphics[width=\textwidth]{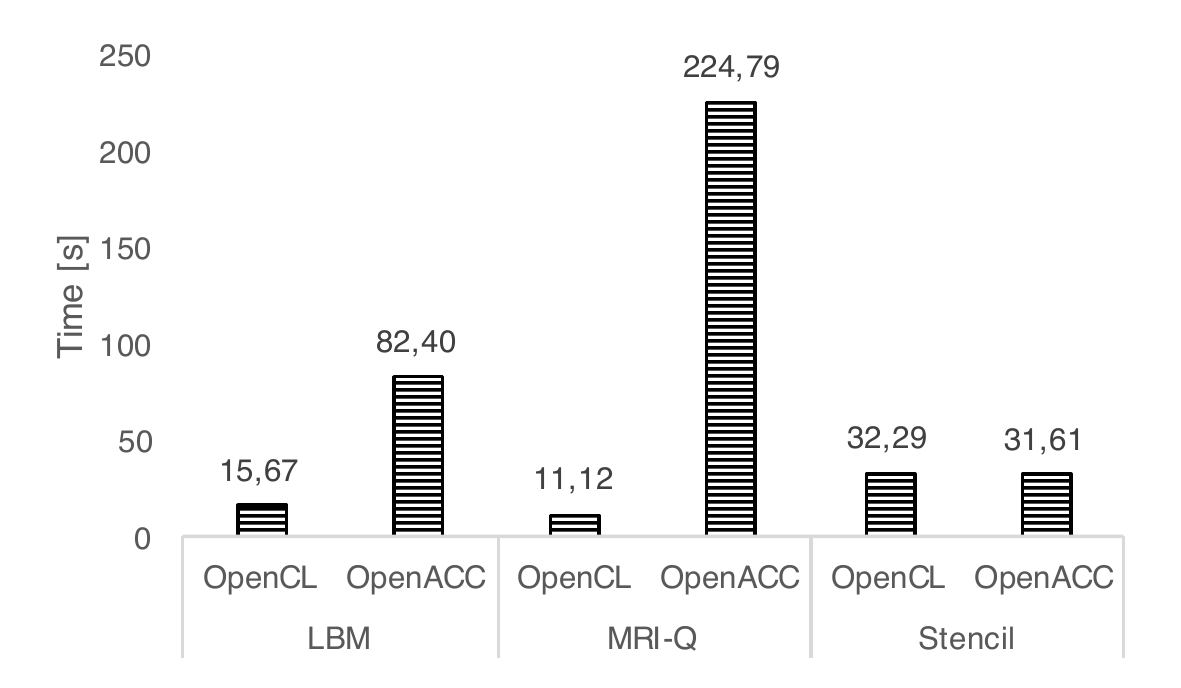}
			\caption{Time}
			\label{fig:time-spec-ida}
		\end{subfigure}
		~ 
		\begin{subfigure}[b]{0.48\textwidth}
			\includegraphics[width=\textwidth]{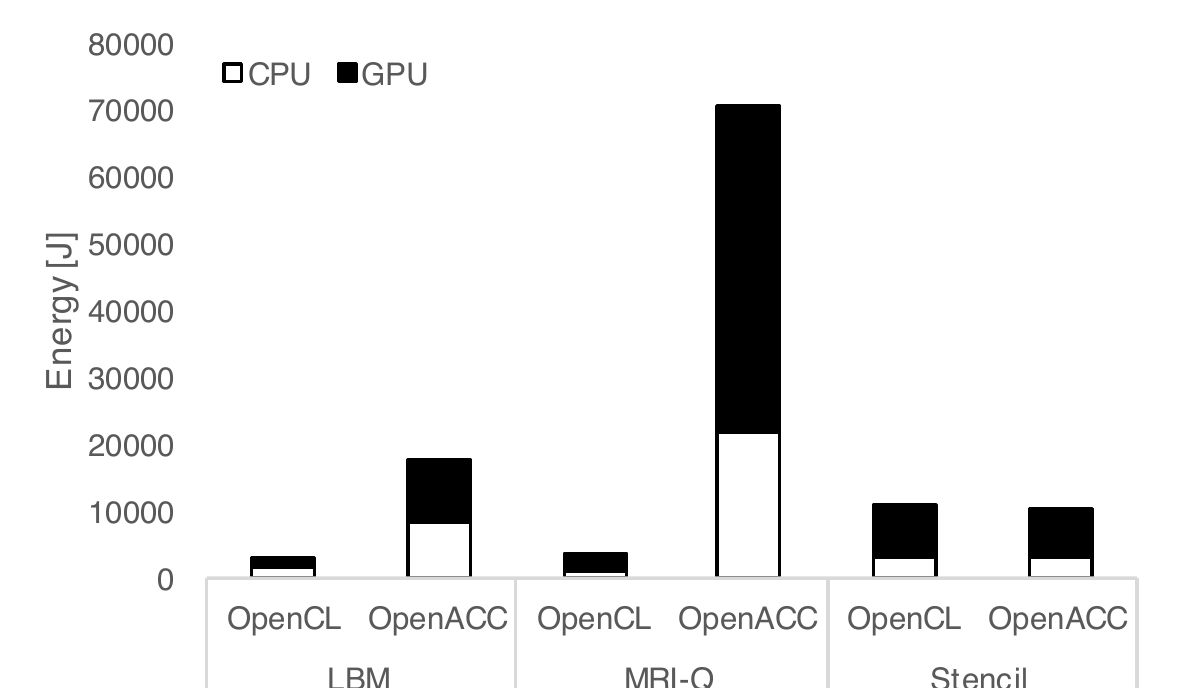}
			\caption{Energy Consumption}
			\label{fig:energy-spec-ida}
		\end{subfigure}
		\caption{A comparison of OpenCL and OpenACC with respect to (a) execution time and (b) energy consumption. Results are obtained using LBM, MRI-Q, and Stencil from the SPEC Accel benchmark suite (Table \ref{table:benchmark-apps}) on the GPU-accelerated system \emph{Ida} (Table \ref{table:system-configurations}).}
		\label{fig:time-energy-spec-ida}
	\end{figure}

	Figures \ref{fig:time-energy-rodinia-ida}, \ref{fig:time-energy-rodinia-emil-ida}, \ref{fig:time-energy-bfs-rodinia-ida}, and \ref{fig:time-energy-spec-ida} depict the execution times and energy consumption for various applications of SPEC Accel and Rodinia benchmark suites on Emil and Ida.
	
	\textit{Result 5: The performance and energy consumption behavior of OpenCL and CUDA are application dependent. For Rodinia benchmark suite, for some applications OpenCL performs better, however there are several applications where CUDA performs better.}
	
	Figure \ref{fig:time-energy-rodinia-ida} depicts the execution time and energy consumption of the OpenCL and CUDA implementations of \emph{NW, B+Tree, GE, Heartwall, Kmeans, LavaMD, SRAD, BP, k-NN, Myocyte, PF}, and \emph{SC} from the Rodinia benchmark suite on the GPU-accelerated system \emph{Ida}. Results show that the OpenCL implementation of seven out of 12 applications, including \emph{NW, GE, LavaMD, SRAD, BP, k-NN}, and \emph{SC} perform better than their corresponding CUDA implementations. 
	
	While for most of the applications, including \emph{B+Tree, GE, SRAD, BP, k-NN, Myocyte}, and \emph{SC}, the performance of OpenCL and CUDA is comparable, for some applications such as \emph{NW} and \emph{LavaMD} there is a large performance difference where OpenCL performs better than CUDA, whereas for \emph{Heartwall, Kmeans,} and \emph{PF} the CUDA implementations perform better than their OpenCL counterparts do.
	
	\textit{Result 6: OpenMP implementation of CFD, HotSpot, and LUD executed on Emil, performs significantly slower than the corresponding OpenCL and CUDA implementation executed on Ida.}
	
	Figure \ref{fig:time-energy-rodinia-emil-ida} depicts the comparison of OpenMP, OpenCL, and CUDA with respect to the execution time and energy consumption using the Rodinia benchmark suite. Figure \ref{fig:time-cfd-hs-lud-rodinia-emil-ida} shows the execution time of \emph{CFD, HotSpot}, and \emph{LUD} applications. The OpenMP implementation of these applications is executed on the Intel Xeon Phi accelerated system \emph{Emil}, whereas the OpenCL and CUDA versions are executed on the GPU accelerated system \emph{Ida}. We may observe that the OpenCL and CUDA versions execution time are comparable, whereas the OpenMP implementation is significantly slower. Please note that the host CPUs of Emil are Intel Xeon E5-2695 v2, whereas the host CPUs of the Ida are of type E5-2650 v4. Furthermore, the Intel Xeon Phi 7120P co-processor on Emil is the first generation of Intel Xeon Phi architecture (known as Knights Corner). 	
	
	\textit{Result 7: OpenMP, OpenCL, and CUDA have comparable performance and energy consumption on Ida.}
	
	Figure \ref{fig:time-energy-bfs-rodinia-ida} shows the comparison of OpenMP, OpenCL, and CUDA with respect to the execution time and energy consumption using the \emph{BFS} from the Rodinia benchmark suite. We may observe comparable performance, where OpenCL is a little bit faster than OpenMP, which is slightly faster than CUDA. According to the results showed in Table \ref{table:programmability} the OpenMP version requires the smallest fraction of code lines (4.86\%) to be written, whereas OpenCL requires 9.07\%, and CUDA 12.5\% of the code lines specific to the corresponding framework.
	
	\textit{Result 8: OpenCL performs better than OpenACC for SPEC Accel benchmark suite on Ida.}
	
	Figure \ref{fig:time-energy-spec-ida} depicts the comparison of OpenCL and OpenACC with respect to execution time and energy consumption using the \emph{LBM, MRI-Q,} and \emph{Stencil} from the SPEC Accel benchmark suite on the GPU-accelerated system \emph{Ida}. 
	We may observe that OpenCL implementation of LBM and MRI-Q is significantly faster than the corresponding OpenACC implementation, whereas for the implementation of Stencil the results are comparable. However, according to the results showed in Table \ref{table:programmability} writing code for OpenCL demands significantly greater effort than writing OpenACC code. 
	
	\section{Related Work}
	\label{sec:rw}
	
	In this section, we highlight examples of research that addresses programming aspects of heterogeneous computing systems and position this paper with respect to the related work.
	
	Su et al. \cite{oclandcuda2012} compare the performance of OpenCL and CUDA using five kernels: Sobel filter, Gaussian filter, median filter, motion estimation, and disparity estimation. The execution time of CUDA-based implementations was 3.8\% -- 5.4\% faster than the OpenCL-based implementations. 
	
	Kessler et al. \cite{kessler2012} study three approaches for programmability and performance portability for heterogeneous computing systems: SkePU skeleton programming library, StarPU runtime system, and Offload C++ language extension. Authors reason about the advantages of each of these approaches and propose how they could be integrated together in the context of PEPPHER project \cite{peppher2011} to achieve programmability and performance portability.
	
	Mittal and Vetter \cite{mittal2015survey}, provide a comprehensive coverage of literature for GPU-accelerated computing systems. In this context, authors survey the literature about runtime systems, algorithms, programming languages, compilers, and applications. 
	
	Li et al. \cite{accandcuda2016} study empirically the programmer productivity in the context of OpenACC with CUDA. The study that involved 28 students at undergraduate and graduate level was performed at Auburn University in 2016. The students received the sequential code of two programs and they had to parallelize them using both CUDA and OpenACC. From 28 students, only a fraction was able to complete properly the assignments using CUDA and OpenACC. The authors conclude that OpenACC enables programmers to shorten the program development time compared to CUDA. However, the programs developed with CUDA execute faster than their OpenACC counterparts do.
	
	This paper complements the related research with an empirical study of four widely used frameworks for programming heterogeneous systems: OpenCL, OpenACC, OpenMP, and CUDA. Using de-facto standard benchmark suites SPEC and Rodinia we study the productivity, performance, and energy consumption. Our measurement tools CodeStat and x-MeterPU that we developed for this study enable us to address systems accelerated with Intel Xeon Phi and GPU. 
	
	\section{Summary}
	\label{sec:summary}
	
	We have presented a study of productivity, performance, and energy for OpenMP, OpenACC, OpenCL, and CUDA. For comparison we used the SPEC Accel and Rodinia benchmark suites on two heterogeneous computing systems: \emph{Emil} that comprises two Intel Xeon Phi E5 processors and one Intel Xeon Phi co-processor, and \emph{Ida} that has two Intel Xeon Phi E5 processors and one GTX Titan X GPU. 
	
	Our major observations include: (1) programming with OpenCL requires significantly more effort than programming with OpenACC for SPEC Accel benchmark suite; (2) programming with OpenCL on average requires about two times more effort than programming with CUDA for Rodinia benchmark suite; (3) the human factor can impact the fraction of code lines to parallelize the code; (4) OpenMP, OpenCL, and CUDA have comparable performance and energy consumption on Ida; and (5) OpenCL performs better than OpenACC for SPEC Accel benchmark suite on Ida; (6) programming with OpenMP requires less effort than programming with OpenCL and CUDA.
	
	Future work will address new architectures of Intel Xeon Phi and NVIDIA GPU. Beside Rodinia and SPEC benchmarks we plan to use real-world applications.
	
\section{Acknowledgment}

This article is based upon work from COST Action IC1406 High-Performance Modelling and Simulation for Big Data Applications (cHiPSet), supported by COST (European Cooperation in Science and Technology.	
	

\end{document}